# Tunable topological edge states in black phosphorus-like Bi(110)

Chen Liu[1*], Shengdan Tao[2*], Guanyong Wang[3], Hongyuan Chen[1], Bing Xia[1], Hao Yang[1,4], Xiaoxue Liu[1,4], Liang Liu[1,4], Yaoyi Li[1,4], Shiyong Wang[1,4], Hao Zheng[1,4], Canhua Liu[1,4], Dandan Guan[1,4#], Yunhao Lu[2,5#], Jin-feng Jia[1,4,6#]

[1]Key Laboratory of Artificial Structures and Quantum Control (Ministry of Education), Shenyang National Laboratory for Materials Science, School of Physics and Astronomy, Shanghai Jiao Tong University, Shanghai 200240, China
[2]School of Physics, Zhejiang University, Hangzhou 310027, China
[3]Shenzhen Institute for Quantum Science and Engineering, Southern University of Science and Technology, Shenzhen 518055, China
[4]Tsung-Dao Lee Institute, Shanghai Jiao Tong University, Shanghai 200240, China
[5]State Key Laboratory of Silicon and Advanced Semiconductor Materials, School of Materials Science & Engineering, Zhejiang University, Hangzhou 310027, China
[6]Department of Physics, Southern University of Science and Technology, Shenzhen 518055, China

Corresponding to: ddguan@sjtu.edu.cn; luyh@zju.edu.cn; jfjia@sjtu.edu.cn

Abstract: We have investigated the structures and electronic properties of ultra-thin Bi(110) films grown on an s-wave superconductor substrate using low-temperature scanning tunneling microscopy and spectroscopy. Remarkably, our experimental results validate the theoretical predictions that the manipulation of Bi(110) surface atom buckling can control the topological phase transition. Notably, we have observed robust unreconstructed edge states at the edges of both 3-bilayer (BL) and 4-BL Bi(110) films, with the 4-BL film displaying stronger edge state intensity and a smaller degree of atomic buckling. First-principle calculations further substantiate these findings, demonstrating a gradual reduction in buckling as the film thickness increases, with average height differences between two Bi atoms of approximately 0.19 Å, 0.10 Å, 0.05 Å, and 0.00 Å for the 1-BL, 2-BL, 3-BL, and 4-BL Bi(110) films, respectively. When Bi films are larger than 2 layers, the system changes from a trivial to a non-trivial phase. This research sets the stage for the controlled realization of topological superconductors through the superconducting proximity effect, providing a significant platform for investigating Majorana zero modes and fabricating quantum devices.

Key words: Scanning tunneling microscopy and spectroscopy, Bi(110) films, topological phase transition, edge states, atomic buckling

Two-dimensional topological insulators (2D TIs), also known as quantum spin Hall insulators, are materials that demonstrate one-dimensional gapless edge states within the bulk insulating gap, exhibiting opposite spin polarizations. These edge states are protected by time-reversal symmetry, making 2D TIs an attractive platform for spintronics and other applications[1-5]. The search for 2D TI materials has garnered significant interest in both theoretical calculations and experimental investigations. The first theoretically predicted 2D TI was HgTe/CdTe quantum wells (QWs) [5], observed in experiments[6] in 2007. Subsequently, several other 2D TI materials have been discovered[7-14], but achieving TIs with high transition temperatures remains a challenging task. Furthermore, realizing a controllable topological phase transition is a challenge for practical device applications.

Freestanding bilayers of group-IV (C, Si, Ge, Sn, and Pb) and V (As, Sb, and Bi) elements in a buckled honeycomb structure have been predicted to exhibit 2D TI properties[9,15-17], which have been confirmed in numerous studies[18-20]. Bismuth (Bi), as the heaviest non-radioactive element, possesses strong spin-orbit coupling, making it a potential candidate for realizing 2D TIs at room temperature. Consequently, Bi films have been extensively studied in various systems[20-24]. Pseudo-cubic Bi(110), a crystallographic orientation of bismuth, has been reported to exhibit nontrivial 2D TI properties, with large and tunable bandgaps determined by the atomic buckling of the Bi(110) films[25]. These films can be classified into three different structures based on the atomic buckling between two nonequivalent Bi atoms within a monolayer (ML): bulk-like structure, black phosphorus (BP)-like structure, and sandwiched structure[25-27]. The BP-like structure in ultrathin Bi(110) films is predicted to exhibit topological properties and has been experimentally studied in various systems with different substrates[25,26,28-33]. However, the structure and electronic properties of ultrathin Bi(110) films are sensitive to electron doping, which is influenced by the substrate. Consequently, reconstructed edges often manifest, making it difficult to discern whether the observed edge states are related to reconstruction or topology. Furthermore, there is currently no experimental work confirming the tunability of Bi(110) films.

In this work, we prepared high-quality ultrathin Bi(110) films on a superconducting substrate and investigated the topography and electronic properties using low-temperature scanning tunneling microscopy and spectroscopy (STM/STS). Our findings revealed the presence of robust edge states at the edges of 3-BL and 4-BL Bi(110) films without reconstruction, with the intensity of edge states significantly stronger in the 4-BL case. These results are consistent with density functional theory (DFT) calculations, suggesting the band topology of Bi(110) is influenced by atomic buckling. Additionally, we observed a gradual decrease in buckling with increasing film thickness and a topological transition between 2-BL and 3-BL films. Hence, the manipulation of surface atomic buckling enables control over the topological phase transition.

The STM image in Figure 1(a) shows typical Bi(110) films, denoted in units of BL (see S1 for details), with their thickness marked by the green numbers. In contrast to prior experimental investigations[26,28,29,31-33], the sample we fabricated exhibits a larger surface area with more linear step edges, and no reconstruction was observed at these step edges (see S1 for more details). Along the red line in Figure 1(a), the height difference between two adjacent layers is about 0.68 nm, as shown in Figure 1(b), consistent with theoretical calculation of the height of 1-BL Bi(110). Figure 1(c) illustrates schematic side and top views of a 1-BL Bi(110) film, where the dark and light blue colors represent the up and down monolayer, respectively. It is important to note that buckling is defined as the height difference (labeled as $h$) between two atoms within the same monolayer. According to our DFT calculations, the unit cell of the Bi(110) film is a rectangle with constants of

4.7 Å ×4.5 Å (depicted as an orange rectangle in Figure 1(c)), with a vertical difference ($h$) of 0.58 Å in each sublayer. Figures 1(d-f) display atomic-resolution STM images of 1-BL, 2-BL and 3-BL Bi(110) films, revealing distinct buckling patterns and atomic arrangements corresponding to each film thickness. Buckling height profiles along the black and red lines, as depicted in Figures 1(g-i), illustrate the transition from a distorted BP structure (DBP) to BP structures, providing crucial visual evidence for the structural transformation and the associated topological phase transition, as supported by previous research results[25].

Theoretical calculations further prove the reliability of the experimental results. DFT calculations (see Method for more computational details) reveal that the stacking configuration between a 5×3 Bi(110) monolayer and a 4√3×4 $NbSe_2$ monolayer exhibits less than 2% lattice mismatches along the armchair and zigzag directions. The most stable structure of multilayer Bi(110) on the $NbSe_2$ substrate by DFT calculations is shown in Figure 2(a). Due to the strong interfacial coupling between Bi(110) and the $NbSe_2$ substrate, the average buckling ($h$) of Bi(110) monolayer rapidly declines from 0.58 Å (in the free-standing state) to 0.19 Å (on the $NbSe_2$ substrate). Additionally, 1-BL Bi(110) films have a rougher surface due to different stacking areas. As Bi(110) films thicken, more Bi-Bi covalent-like quasi-bands appear rather than pure van der Waals (vdW) interactions (Figure 2(a) and see S2 for details), and the Bi(110) surface becomes smoother. To provide a clearer comparison of buckling in different Bi(110) layers, the average buckling ($h$) of surface Bi atoms has been calculated, as depicted in Figure 2(b), revealing a consistent trend of decreasing buckling with increasing film thickness. Both the STM and DFT results confirm the gradual decrease in buckling as the film thickness increases. Experimentally, the measured $h$ values are 0.19 Å, 0.09 Å and 0.07 Å for 1-BL, 2-BL and 3-BL Bi(110) films, respectively, which reasonably agree with the DFT-calculated values of 0.19 Å, 0.10 Å, and 0.05 Å. Notably, the 4-BL Bi(110) film exhibits an almost flat buckling profile. This demonstrate that buckling is closely related to the thickness of Bi(110) film.

To numerically depict the lateral electronic modulations, we investigated the local density of states (LDOS) of 1-BL, 2-BL, 3-BL, and 4-BL Bi(110) films, where different colors represent different layers. As illustrated in Figure 2(c), the 1-BL, 3-BL, and 4-BL Bi(110) films exhibit moderate LDOS values at Fermi level ($E_F$), while the 2-BL Bi(110) films exhibit strong peaks near $E_F$. Considering that the experimentally measured electronic states mainly come from 1-BL Bi on the surface, we calculated the LDOS of the upper 1-BL Bi, as shown in Figure 2(d). The DFT results closely resemble the experimental measurement. Specifically, for 2-BL Bi(110), such a high LDOS peak at $E_F$ is approximately 0.1 eV width. In other cases, there are fewer electronic states located at $E_F$. The thickness-dependent modulation of LDOS around $E_F$ further proves that the buckling effect changes the distribution of electronic states of Bi(110).

It should be noted that Lu et al.[25] have investigated topological properties determined by atomic buckling in bilayer Bi(110). By using the generalized-gradient approximation (GGA) instead of local density approximation (LDA) for the exchange and correlation functional, we have calculated the band structures and density of states of bilayer Bi(110) with different buckling ($h$) and the results are shown in S3. Similarly, as buckling decreases (from 0.58 Å to 0.00 Å), the bilayer Bi(110) transforms from topologically trivial to nontrivial with an inverted bandgap. The difference is the appearance of gapless edge states occur at $h = 0.09$ Å, slightly below the 0.1 Å[25], accompanied by an increased density of electron states around $E_F$. According to the relationship between topological properties and atomic buckling discussed above, we can reasonably infer to a conclusion that 1-BL Bi(110) ($h$=0.19 Å) and 2-BL Bi(110) ($h$=0.09 Å) do not possess edge states, while 3-BL Bi(110) ($h$=0.07 Å) and 4-BL Bi(110) (almost zero $h$) are expected to have edge states in STM images. This hypothesis will be further discussed in detail. Interestingly, the buckling of 2-BL Bi(110) films ($h$ = 0.10 Å) is close to the topological phase transition point in bilayer Bi(110) at $h$ = 0.09 Å, which may contribute to the presence of additional electronic states located at $E_F$. Moreover, considering that the moiré pattern may cause band flattening[34-36]. This is a reasonable reason why a high peak appears at $E_F$ in the 2-BL Bi(110) films. Based on the aforementioned results and analysis, we could observe a consistent relationship between the thickness, buckling and topology of Bi(110).

To demonstrate the potential topological phase transition induced by the structural transformation, we conducted a thorough experimental study in the system. Figures 3(a-c) display the STM images of Bi(110) films with thicknesses of 2-BL, 3-BL, and 4-BL, respectively. Meanwhile, Figures 3(d-f) present the profiles of the LDOS intensity obtained along the red arrows. These arrows are nearly perpendicular to the edge, as shown in Figures 3(a-c), respectively. A discernible edge state emerges close to the edge of the 3-BL Bi(110) at +310 mV, spanning a spatial range of roughly 1 nm. In parallel, a significant edge state is noted near the edge of the 4-BL Bi(110) at both -200 mV and +220 mV, covering a spatial range of roughly 5 nm. The observations indicating these phenomena are denoted by the presence of red dashed circles in Figures 3(e) and (f). In contrast, there is no edge state signal near the edge of the 2-BL Bi(110), which is confirmed by the LDOS intensity profile in Figure 3(d) and the LDOS map (+300 mV) in Figure 3(g). The enduring presence of robust edge states in 3-BL and 4-BL Bi(110) is depicted in the LDOS maps at +311 mV and -260 mV, respectively, as shown in Figures 3(h) and (i) (see S4 for details). To further describe the edge position and strength of different layers, Figures 3(j-l) show the LDOS intensity profiles along black (in Figures 3(a-c)) and red(in Figures 3(g-i)) lines, where the black line corresponds to topography and red line corresponds to LDOS map. Notably, the edge state signal in the 4-BL Bi(110) is much stronger and remains unaffected by variations in the morphology of the step edges, which further confirms that the edge states are topologically protected. Based on our study, it's possible to realize controlled topological superconductors experimentally due to the superconducting proximity effect (see S5 for details).

To gain further insight into the topological nature of ultrathin Bi(110) films by considering the bulk-boundary, we investigated the edge states of a bilayer Bi(110) nanoribbon due to the buckling variation. The 3-BL Bi(110) film exhibits a topological nontrivial behavior with the appearance of topological edge states, as its buckling is below the topological transition point. For 4-BL Bi(110), there is almost no buckling, showing strong topological boundary performance. We calculated band structures of bilayer Bi(110) nanoribbons with 0.05 Å and 0 Å buckling, which correspond exactly to the average buckling of surface Bi atoms in 3-BL and 4-BL Bi(110) on NbSe$_2$ by DFT. As shown in Fig.4, both edge bands of the two different buckling cases are located inside the bulk bandgap, exhibiting linear crossing at Y of the 1D BZ, and merge to the bulk states near Γ. But the degeneracy of edge states increase with buckle decreasing. For example, near the Y point (blue dashed line), the edge states at $h$=0 Å buckling has a higher degeneracy than $h$=0.05 Å buckling (see Fig. 4(a,b)), which also corresponds to a stronger edge states intensity (see Fig.4(c)). To unveil the edge state intensity to buckling, we performed the spatial distribution of edge states, as shown in Figures S6(a) and (b). For a buckling of 0.05 Å, the range of the *k*-dependent wave function extending ~1 nm (Figure S6(a)) has the good consistency to the STM images (Figure 3(e)). For a buckling of 0.00 Å, the range of the wave function extending 5 nm (Figure S6(b)) also accords quite well with the STM images (Figure 3(f)). Comparing the two cases, the topological edge states are more robust for a buckling of 0.00 Å. Our DFT results about edge states at different buckling values are consistent with the STM images, further confirming the buckling-dependent topological edge states. It is clear that the flatness of Bi(110) film is vital in determining the topological edge states. Moreover, the agreement between experimental measurement and calculation results of 1-BL nanoribbon further proves that the topological edge states of ultrathin Bi(110) films are robust against substrate effect.

In summary, we have successfully fabricated high-quality Bi(110) films on NbSe$_2$ substrate and observed robust edge states in 3-BL and 4-BL Bi(110), with significantly stronger intensity in the latter. Our experiments and theoretical calculations suggest that the degree of buckling gradually decreases as the film thickness increases, while the gapless edge states exhibit enhanced strength. Notably, this study represents the first instance where the control of the topological phase transition is achieved through the manipulation of surface atomic buckling. This work opens up new possibilities for realizing controlled topological superconductors by leveraging the superconducting proximity effect. And it provides a promising platform for investigating Majorana zero modes and fabricating quantum devices.

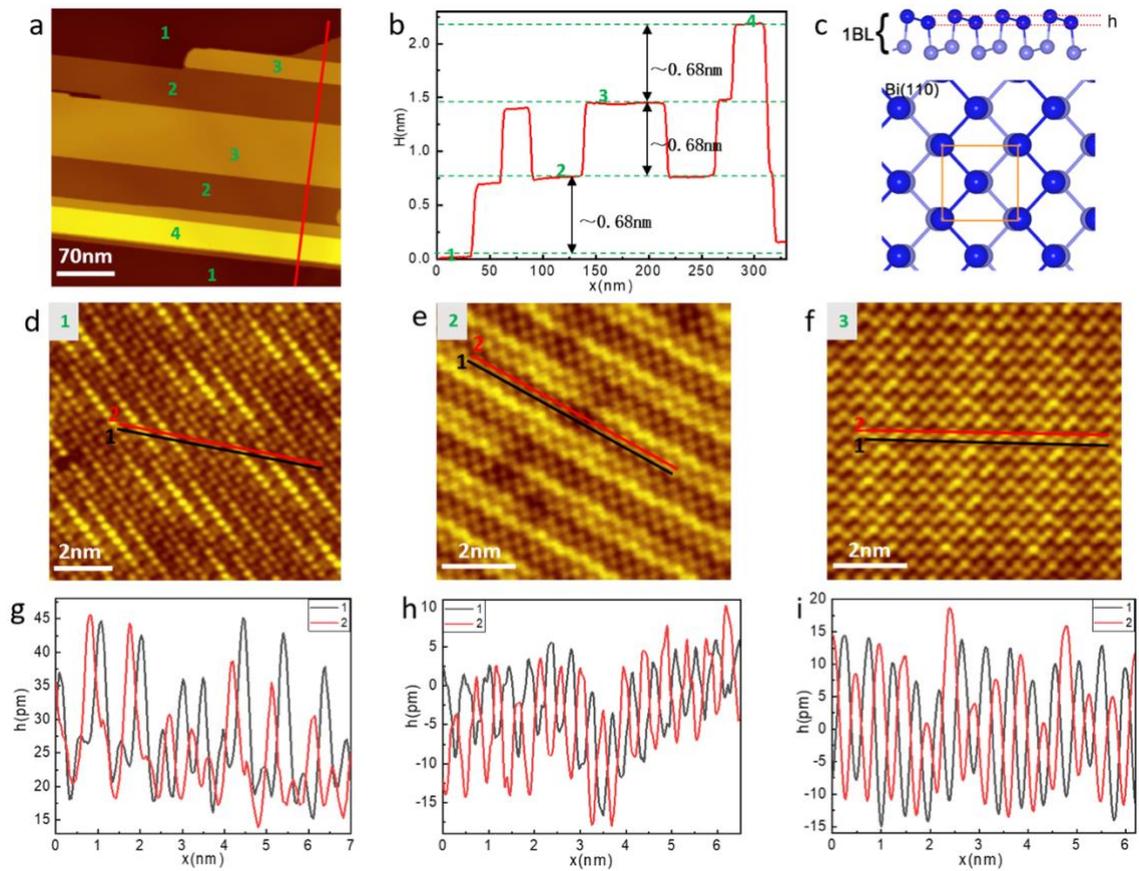

**FIG. 1.** (a) STM image of Bi(110) films formed on NbSe$_2$ substrate. (b) The height profile along red line in (a). (c) Schematic side and top views of 1-BL Bi(110) film, where buckling is defined in terms of the height difference between two atoms of the same monolayer. The up and down monolayer are marked by dark blue and light blue balls, respectively. (d, e, f) Atomic-resolution STM images of 1-BL, 2-BL and 3-BL Bi(110) films, respectively. (g, h, i) The buckling height profiles of different BLs for the black and red lines 1 and 2 in (d), (e) and (f), respectively.

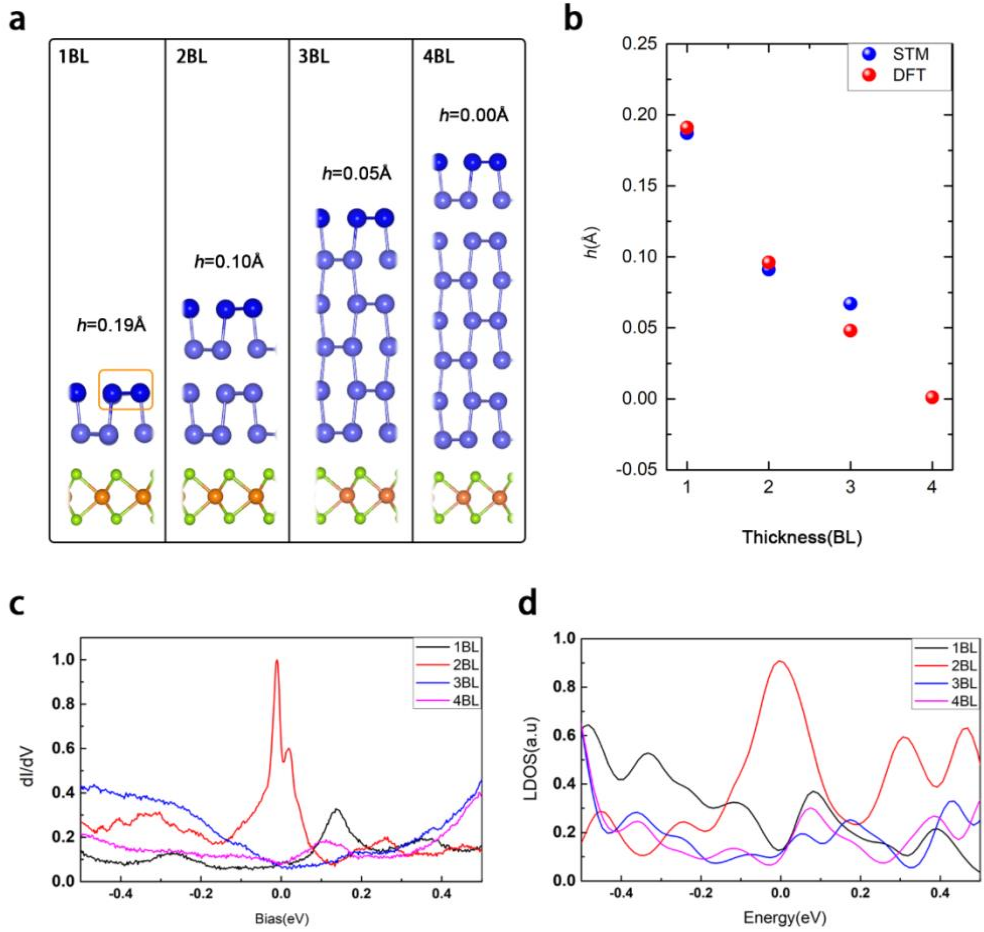

**FIG. 2.** (a) The average buckling $h$ of DFT calculation for 1-BL, 2-BL, 3-BL and 4-BL Bi(110) films. (b) The average buckling $h$ contrast between STM and DFT. (c) LDOS of 1-BL, 2-BL, 3-BL and 4-BL Bi(110) films on NbSe$_2$ substrate by STM. (d) LDOS of the upper 1-BL for 1-BL to 4-BL Bi(110) films on NbSe$_2$ substrate by DFT calculation.

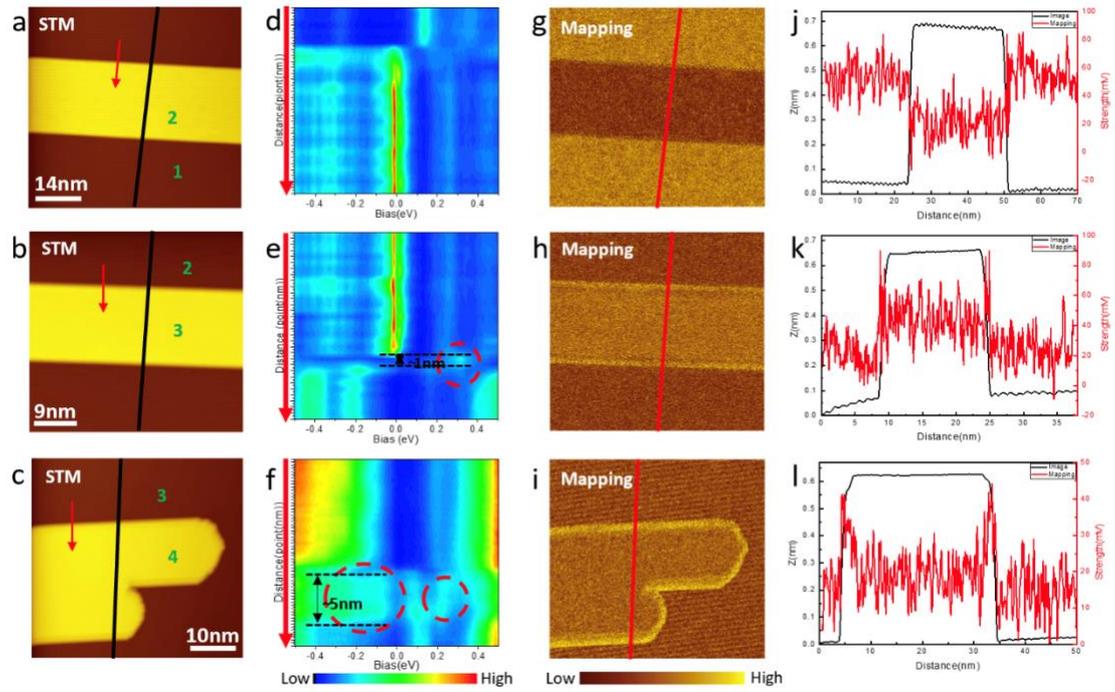

**FIG. 3.** (a, b, c) STM images of Bi(110) film with different thickness, and the thickness information are marked by green numbers. (d, e, f) LDOS intensity profiles measured along the red arrows in (a), (b) and (c), respectively. The red arrow indicates the measurement direction, and the red dotted circle indicates the region where the edge state is located. (g), (h) and (i) LDOS maps at +300 mV, +311 mV and -260 mV correspond to (a), (b) and (c), respectively. (j, k, l) The LDOS intensity profiles along black and red lines in topography(a, b, c) and LDOS map(g, h, i).

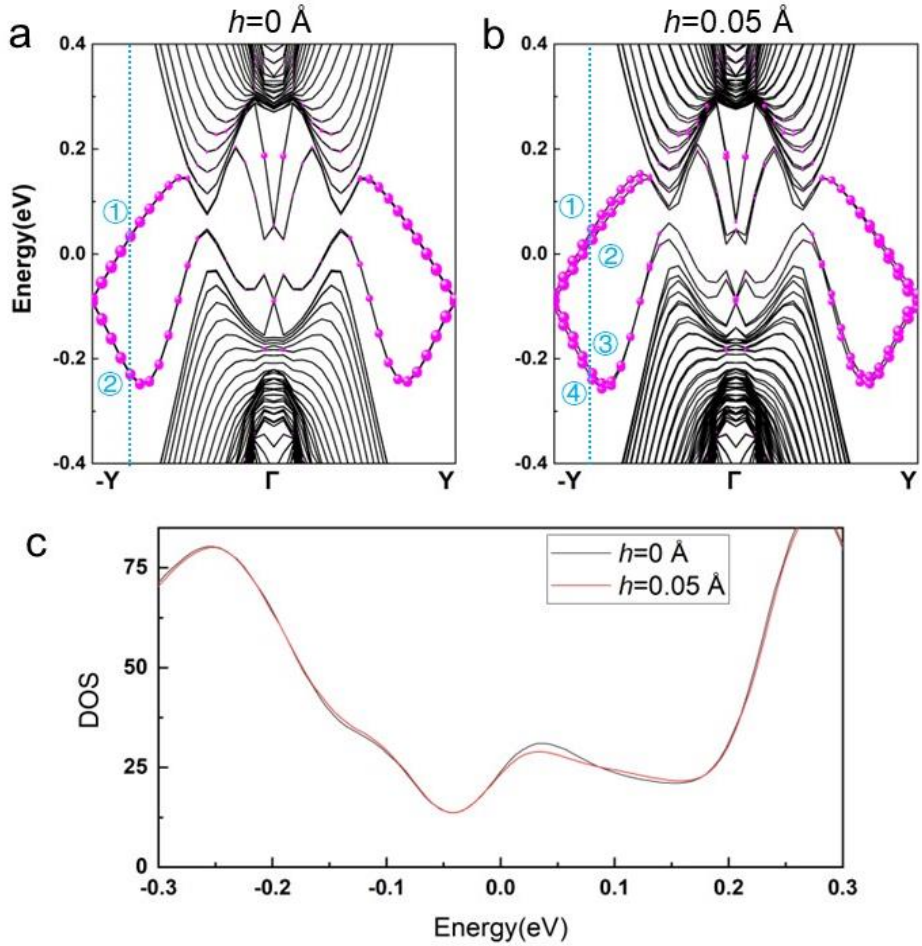

**FIG. 4.** (a, b) Band structures and (c) DOS of the ribbon 1BL Bi(110) with 0 Å and 0.05 Å buckling by DFT calculation.

# METHODS

**MBE Growth**

The 2H-NbSe$_2$ substrate is degassed at about 500K until the vacuum degree is restored to $5.0 \times 10^{-10}$ Torr, then it is cleaved in ultrahigh vacuum (2 Torr) at about 298 K to obtain a micron level flat NbSe$_2$ (0001) cleavage plane. We deposit high-purity Bi (99.999%) atoms from a standard Knudsen cell onto the prepared substrate surface where the temperature is about 373 K, and the deposition rate is about 0.003 ML/s. After deposition, slowly cooling the system to room temperature.

**STM/ STS Measurements**

The experiments are performed on a commercial Unisoku ultrahigh vacuum MBE-STM system (1300), and data are measured at the liquid helium temperature of about 4 K. Cleaned W tips are checked on Ag islands before the STM and STS measurements. All topographic images are obtained in a constant-current mode, while the tunneling dI/dV is recorded by standard lock-in technique, with the feedback off.

Computational Details

Our first-principles calculations were carried out within the density functional theory based on the projected augmented wave (PAW)[37,38] pseudopotentials as implemented in the Vienna Ab-initio Simulation Package (VASP)[39]. The generalized-gradient approximation (GGA) in the form of Perdew-Burke-Ernzerhof (PBE)[40] functional was used. The cutoff of the plane wave basis was set to guarantee that the absolute energies are converged to a few meV. The DFT-D3 method with Becke-Johnson damping[41] was employed to incorporate the effects of vdW interactions. The vacuum region was set larger than 15 Å to minimize artificial interactions. Firstly, we consider the stacking order of the Bi(110) in multiple-layer structures. We change relative position by moving and flipping one of the layers. All possible stable structures are listed in S3. Among them, AA stacking has the lowest energy, which means it is most likely to show up multiple-layer Bi(110). According to this result, the following calculation are based on AA stacking. Secondly, as shown in S7, we test the buckling of Bi(110) in different layer numbers without substrate. It is clearly seen that buckling of Bi(110) decreases with the increase of layers. Next, we take the effect of $NbSe_2$ substrate into consideration. In order to minimize the lattice mismatch and reduce the calculation cost. The unit cell of $Bi/NbSe_2$ heterostructure comprises 5 × 3 Bi(110) films and a 4 √3 × 4 $NbSe_2$ substrate. The lattice mismatch along the armchair (x) and zigzag (y) directions are both below the 2%. In order to further explain the between edge states and buckling, we calculate the 1-BL Bi(110) nanoribbon. Because no reconstruction was observed experimentally, we only consider the odd and even numbers for each sub-layer, and Spin-orbital coupling (SOC) is included.

**Acknowledgements**

We thank the Ministry of Science and Technology of China (Grants No. 2019YFA0308600, 2020YFA0309000), NSFC(Grants No. 92365302, No. 22325203, No. 92265105, 92065201, No. 12074247, No. 12174252), the Strategic Priority Research Program of Chinese Academy of Sciences (Grant No. XDB28000000) and the Science and Technology Commission of Shanghai Municipality (Grants No. 2019SHZDZX01, No. 19JC1412701, No. 20QA1405100) for partial support. We thank the financial support from Innovation program for Quantum Science and Technology (Grant No. 2021ZD0302500).

Supplemental Material

# Tunable topological edge states in black phosphorus-like Bi(110)

Chen Liu[1*], Shengdan Tao[2*], Guanyong Wang[3], Hongyuan Chen[1], Bing Xia[1], Hao Yang[1,4], Xiaoxue Liu[1,4], Liang Liu[1,4], Yaoyi Li[1,4], Shiyong Wang[1,4], Hao Zheng[1,4], Canhua Liu[1,4], Dandan Guan[1,4#], Yunhao Lu[2,5#], Jin-feng Jia[1,4,6#]

[1]Key Laboratory of Artificial Structures and Quantum Control (Ministry of Education), Shenyang National Laboratory for Materials Science, School of Physics and Astronomy, Shanghai Jiao Tong University, Shanghai 200240, China
[2]School of Physics, Zhejiang University, Hangzhou 310027, China
[3]Shenzhen Institute for Quantum Science and Engineering, Southern University of Science and Technology, Shenzhen 518055, China
[4]Tsung-Dao Lee Institute, Shanghai Jiao Tong University, Shanghai 200240, China
[5]State Key Laboratory of Silicon and Advanced Semiconductor Materials, School of Materials Science & Engineering, Zhejiang University, Hangzhou 310027, China
[6]Department of Physics, Southern University of Science and Technology, Shenzhen 518055, China

## Contents

**S1. Sample preparation and STM measurement**
**S2. The influence of the substrate on the Bi(110) thin films**
**S3. A discussion of how multiple-layer Bi(110) are stacked**
**S4. The band topology of Bi(110) thin films controlled by the buckling**
**S5. The superconducting properties in Bi(110) thin films**
**S6. The the spatial distribution of edge states of the ribbon 1BL Bi(110) with 0 Å and 0.05 Å buckling**
**S7. The influence of layer numbers on atomic buckling**

**S1. Sample preparation and STM measurement**
All experiments were studied by a commercial UHV molecular-beam-epitaxy STM system at a base pressure less than Torr, and all experimental data were measured at the liquid helium temperature of about 4 K.

The substrate used in the experiment is 2H-NbSe$_2$, a s-wave superconductor[1-3]. In experiment, the sample before cleavage is degassed at a temperature higher than the growth of Bi(110)($\sim$ 500 K)until the vacuum degree is restored to about $5.0\times 10^{-10}$ Torr, and the sample can be cleaved to obtain a micron level flat NbSe$_2$ (0001) cleavage plane.

Then, we deposit high-purity Bi(99.999%) atoms on the prepared NbSe$_2$ substrate to obtain Bi(110) films. Here we use bilayers(BLs) as unit of the coverage, where 1 BL means that all lattice sites are completely covered with one layer Bi film. The deposition time is about 11 min. The deposition rate is about 0.003 ML/s. The deposition temperature is about 373K. After deposition, we slowly cool the system to room temperature, and successfully prepared 2 BL Bi(110) films on NbSe$_2$ substrate. The topography of the 2 BL films are shown in Figs. S1(a) and (b), where the surface area of Bi(110) ultrathin films are clean and large, and the step edges are very uniform and straight. Fig. S1(c) shows the atomic-resolution STM image of edge between second(2-BL) and third(3-BL) layers of Bi(110), without any reconstruction observed at the step edges.

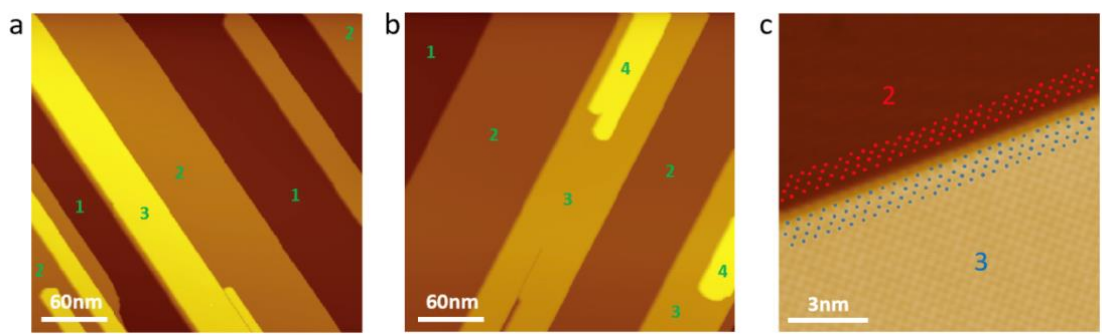

**FIG. S1. STM images of Bi(110) on NbSe$_2$.** (a, b) The topography of 2 BL Bi(110) on NbSe$_2$ substrate. (c) Atomic-resolution STM image of edge between second(2-BL) and third(3-BL) layers of Bi(110), where the red dots represent the second layer(2-BL) of Bi atoms and the blue dots represent the third layer(3-BL) of Bi atoms.

## S2. The influence of the substrate on the Bi(110) thin films.

The Fig.S2 shows the charge density difference plot between the topmost surface layer (BL Bi(110)) and the layer below for 1-BL, 2-BL, and 3-BL Bi(110) structures. It can be observed that when growing 1BL-Bi(110) on the NbSe$_2$ substrate, there is a significant interlayer charge transfer between Bi(110) and the substrate. As the thickness increases to 2-BL and beyond, the charge transfer remains relatively constant with increasing Bi(110) thickness. The results, as shown in Fig. S2(b), indicate that by optimizing the structure under fixed interlayer distances of 3.00 Å, 4.00 Å, and 5.00 Å with the NbSe$_2$ substrate, the average buckling of surface Bi atoms increases, measuring $h$=0.05 Å, 0.27 Å, and 0.31 Å, respectively. This trend once again reflects the strong interaction between the substrate and Bi(110), but as the thickness increases, the influence of the substrate on the surface Bi gradually diminishes. We could observe a consistent relationship between the substrate, thickness and buckling of Bi(110). The substrate and the layer thickness are the direct causes, while the atomic buckling is the fundamental reason of topological phase transition.

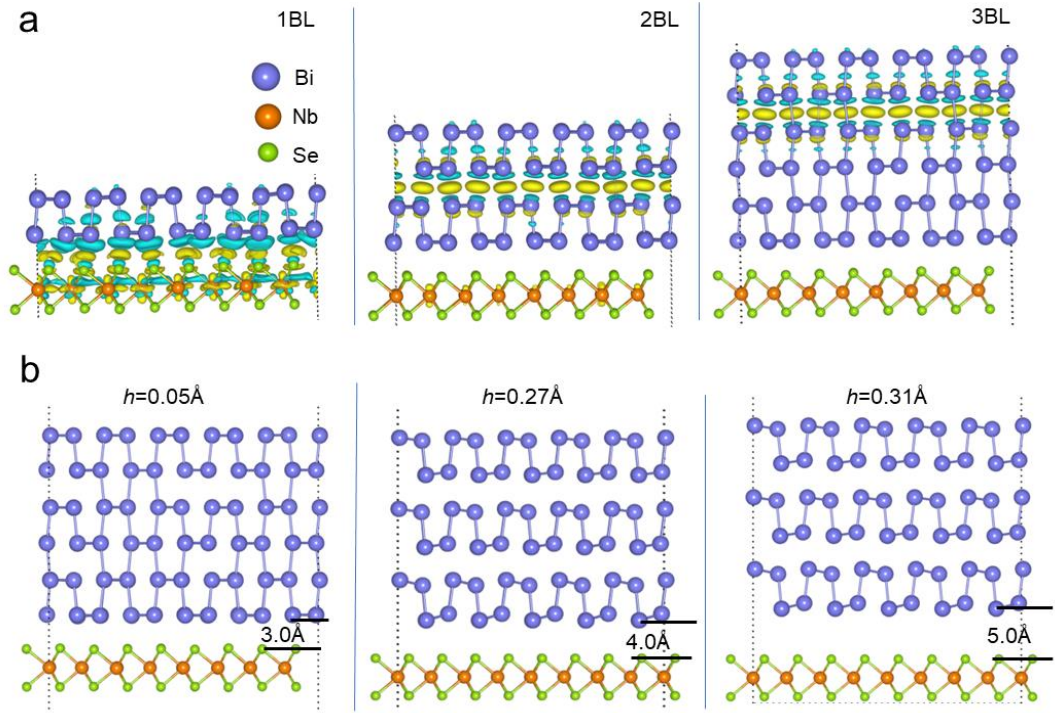

**FIG. S2. Bi(110) thin films on NbSe$_2$ substrate.** (a) The charge density difference plots between the surface BL Bi(110) and the BL Bi(110) below in the 1-BL, 2-BL, and 3-BL Bi(110) structures. Blue indicates charge depletion, while yellow indicates charge accumulation. The isosurface is set to 0.0012 e/Bohr$^3$. (b) The buckling of 3-BL Bi(110) structure as a function of the distance from the NbSe$_2$ substrate.

## S3. A discussion of how multiple-layer Bi(110) are stacked

We change relative position by moving and flipping one of the layers. All possible stable structures are listed in Fig.S3(a). Fig.S3(b) shows the 2-BL Bi(110) unit cell energy of each structure which calculated by DFT. Among them, AA stacking has the lowest energy, which means it is most likely to show up multiple-layer Bi(110).

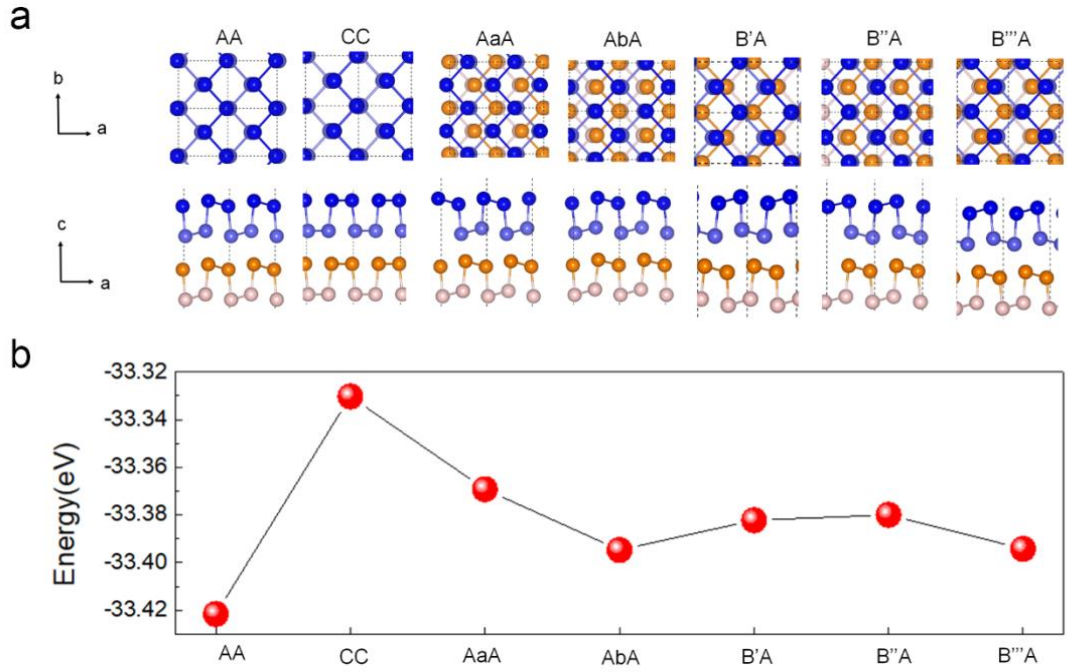

**FIG. S3. The energy of different stacking orders in 2-BL Bi(110).** (a) Top and side views of different stacking orders in 2-BL Bi(110), and the corresponding energies of these configuration are shown in (b).

## S4. The band topology of Bi(110) thin films controlled by the buckling

Fig. S4(a) show the band structures and density of states of 1-BL Bi(110) from free relaxed structure to no buckling structure with SOC considering. Fig. S4(b) shows the high symmetry points of the two-dimensional Brillouin zone. It is obvious that the band gap between Y and Γ points decreases with atomic buckling $h$ and then opens again accompanied by topological transformation. Fig. S4(c) is the graph of the band gap ($\Delta$) with atomic buckling $h$. The gapless edge states appear when $h$ is below the 0.09Å, and Bi(110) become non-trivial. In particular, at this point, there are more electron states around Fermi level.

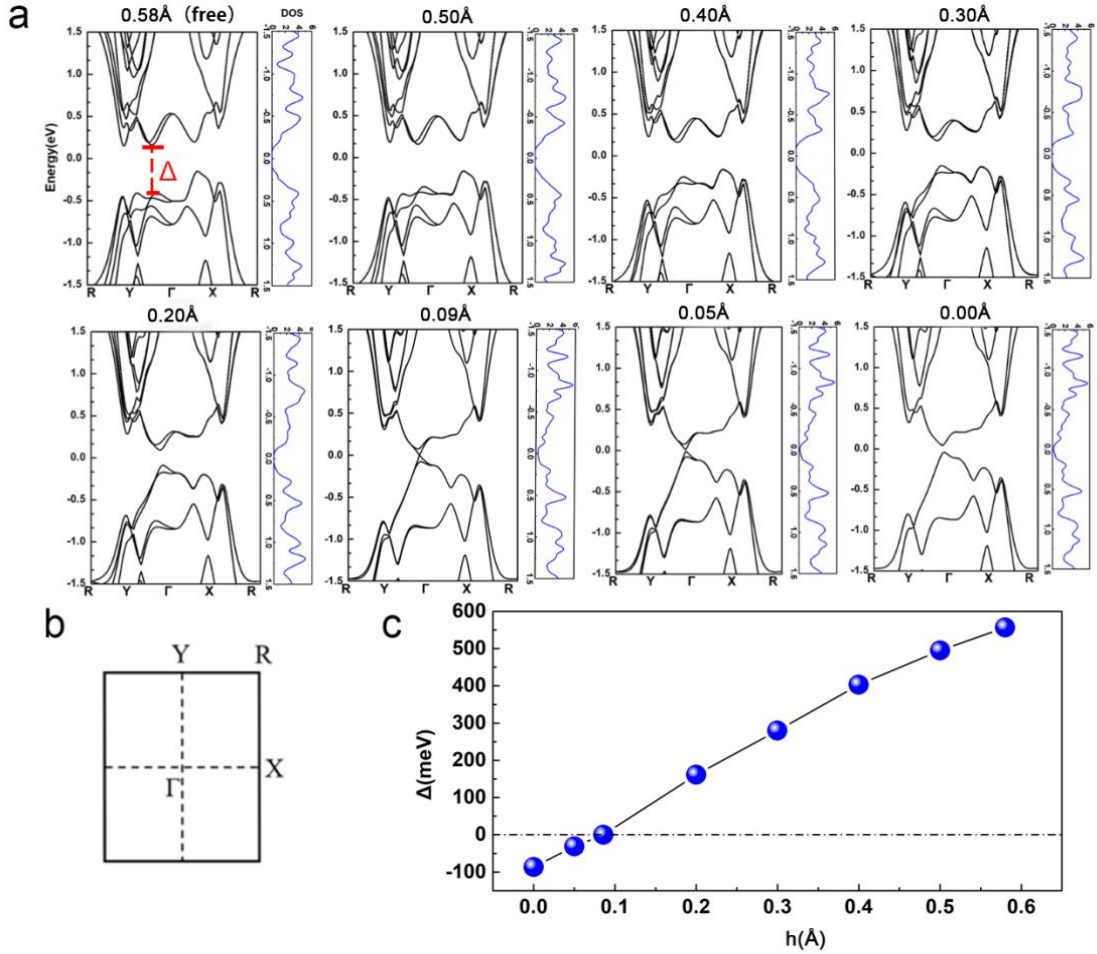

**FIG. S4. The band topology of Bi(110) thin films controlled by the buckling.** (a) The band structures and density of states of 1-BL Bi(110) with different buckling and the high symmetry points of the two-dimensional Brillouin zone are showed in (b). (c) The figure of the band gap ($\Delta$) with atomic buckling $h$.

## S5. The superconducting properties in Bi(110) thin films

$NbSe_2$, a typcal S-wave superconductor, is a good candidate for substrate. Due to the good superconductivity proximity effect, it provides a platform for exploring Majorana zero mode, which has been confirmed in many experiments[4-7].

Fig. S5(a) shows the typical Bi(110) films with the thickness marked by the green numbers. In our study, Bi(110) films formed on the superconducting substrate($NbSe_2$). According to superconducting proximity effect, the superconducting properties are observed in Bi(110) films as shown in Fig. S5(b), where superconducting gap is about 300 meV. The STS spectrum is measured at solid red circle in Fig. S5(a), and measuring temperature is about 400 mk.

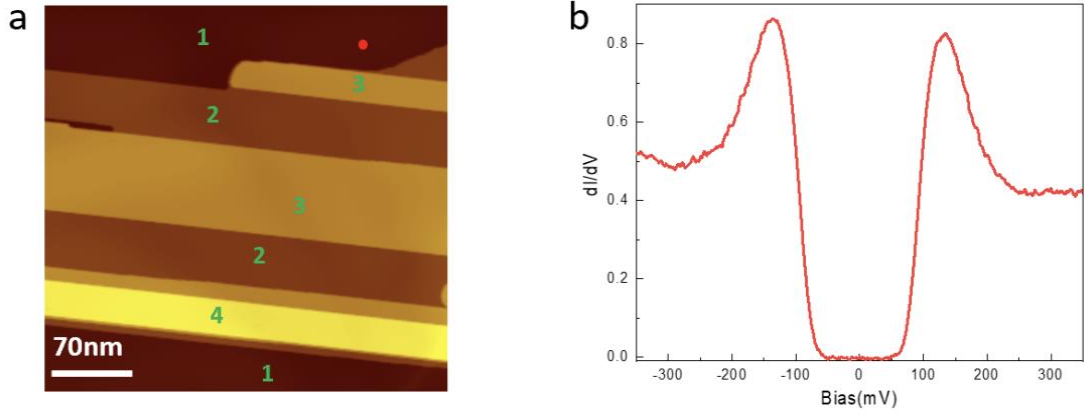

**FIG. S5. The superconducting properties in Bi(110) thin films.** (a) STM image of Bi(110) films formed on NbSe$_2$ substrate. (b) STS spectrum at solid red circle in (a).

## S6. The the spatial distribution of edge states of the ribbon 1BL Bi(110) with 0 Å and 0.05 Å buckling

To unveil the edge state intensity to buckling, we performed the spatial distribution of edge states, as shown in Figures S6(a) and (b). For a buckling of 0.05 Å, the range of the *k*-dependent wave function extending ~1 nm (Figure S6(a)) has the good consistency to the STM images (Figure 3(e)). For a buckling of 0.00 Å, the range of the wave function extending 5 nm (Figure S6(b)) also accords quite well with the STM images (Figure 3(f)).

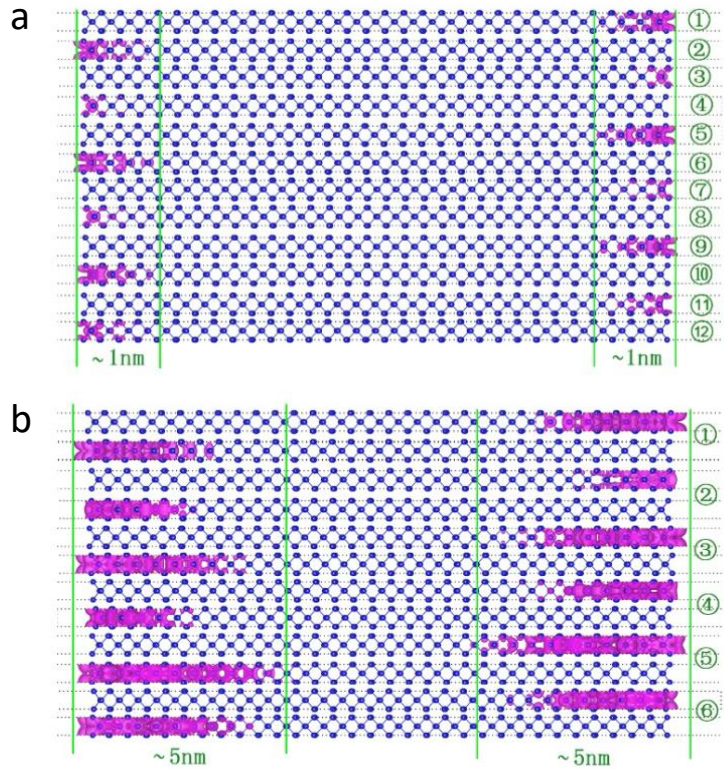

FIG. S6. (a, b) Calculated real space charge density distribution of the edge states at different k-points as marked by the magenta circles in Fig. 4(a, b), respectively. The isosurface is $4\times10^{-5}$e/Bohr$^3$.

## S7. The influence of layer numbers on atomic buckling

We test the buckling of Bi(110) in different layer numbers without substrate, the result is shown in Fig.S7. The *h* calculated is 0.59 Å, 0.56 Å, 0.46 Å and 0.39 Å for 1-BL, 2-BL, 3-BL and 4-BL Bi(110) films. It is clearly seen that buckling of Bi(110) decreases with the increase of layers. By comparing the results with those on $NbSe_2$ substrate, the existence of substrate makes Bi(110) flatter.

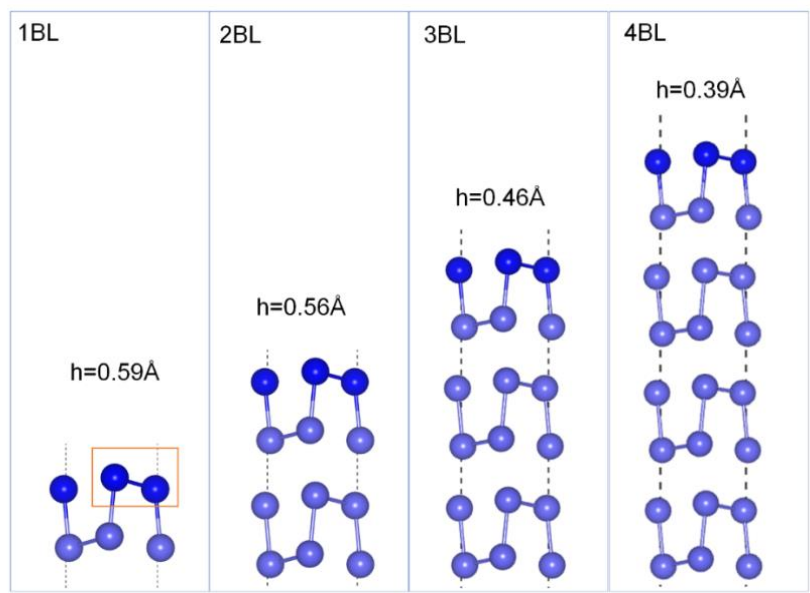

**FIG. S7. Buckling of Bi(110) in different layer numbers without $NbSe_2$ substrate.**